\documentclass[11pt,twoside]{article}
\usepackage{asp2010}

\resetcounters

\begin{document}

\title{Dissecting Kinematics and Stellar Populations of
  Counter-Rotating Galaxies with 2-Dimensional Spectroscopy}
\author{L.~Coccato,$^{1,2}$ L. Morelli,$^{3,4}$ A.  Pizzella,$^3$
  E. M. Corsini,$^{3,4}$ E. Dalla Bont\`a,$^{3,4}$ and L. M. Buson$^4$}
\affil{$^1$European Southern Observatory, Garching bei M\"unchen, Germany}
\affil{$^2$Institute of Cosmology and Gravitation, 
    University of Portsmouth, Portsmouth, UK}
\affil{$^3$Dipartimento di Fisica ed Astronomia `Galileo Galilei',
    Universit\`a di Padova, Padova, Italy}
\affil{$^4$INAF-Osservatorio Astronomico di Padova, Padova, Italy}

\begin{abstract}
We present a spectral decomposition technique and its applications to
a sample of galaxies hosting large-scale counter-rotating stellar
disks. Our spectral decomposition technique allows to separate and
measure the kinematics and the properties of the stellar populations
of both the two counter-rotating disks in the observed galaxies at the
same time. Our results provide new insights on the epoch and mechanism
of formation of these galaxies.
\end{abstract}

\section{Introduction}

Some galaxies, despite of their simple morphologies, can reveal to be
very complex sistems from a kinematic point of view. Clear examples are
objects with stars and/or gas in orthogonal rotation with respect
the rest of the galaxy (e.g., \citealt{Corsini+03, Coccato+04,
  Corsini+12}), or with a warped inner structure (e.g.,
\citealt{Coccato+07}).  Among these galaxies with multi-spin
components, {\em counter-rotating galaxies\/} are those that host two
components that rotate in opposite directions from each other. These
peculiar objects have been observed from in all morphological
types. They are classed by the nature (stars vs. stars, stars vs. gas,
gas vs. gas) and size (counter-ro\-ta\-ting cores, rings, and disks)
of the counter-rotating components (see \citealt{Bertola+99}, for a
review).  The subject of this work is the peculiar case of galaxies
hosting two counter-rotating stellar disks of comparable size.  The
prototype of this class of objects is the famous E7/S0 galaxy
NGC~4550, whose counter-rotating nature was first discovered by
\citet{Rubin+92}.

The study of the kinematics and the stellar population of the two
counter-rotating components are the key to understand the formation
mechanisms of these objects. This task is complicated by the fact that
the two components are co-spatial, and therefore we observe their
combined contribution at each position on the sky. Thus, it is
necessary to separate their contribution to the total observed galaxy
spectrum and measure the properties and kinematics of their stellar
populations separately. To this aim we developed a spectroscopic
decomposition technique and successfully applied to three spiral
galaxies with counter-rotating stellar disks \citep{Coccato+11,
  Coccato+13b, Coccato+13}.

\begin{figure}[t!]
\centering
\includegraphics[width=13cm]{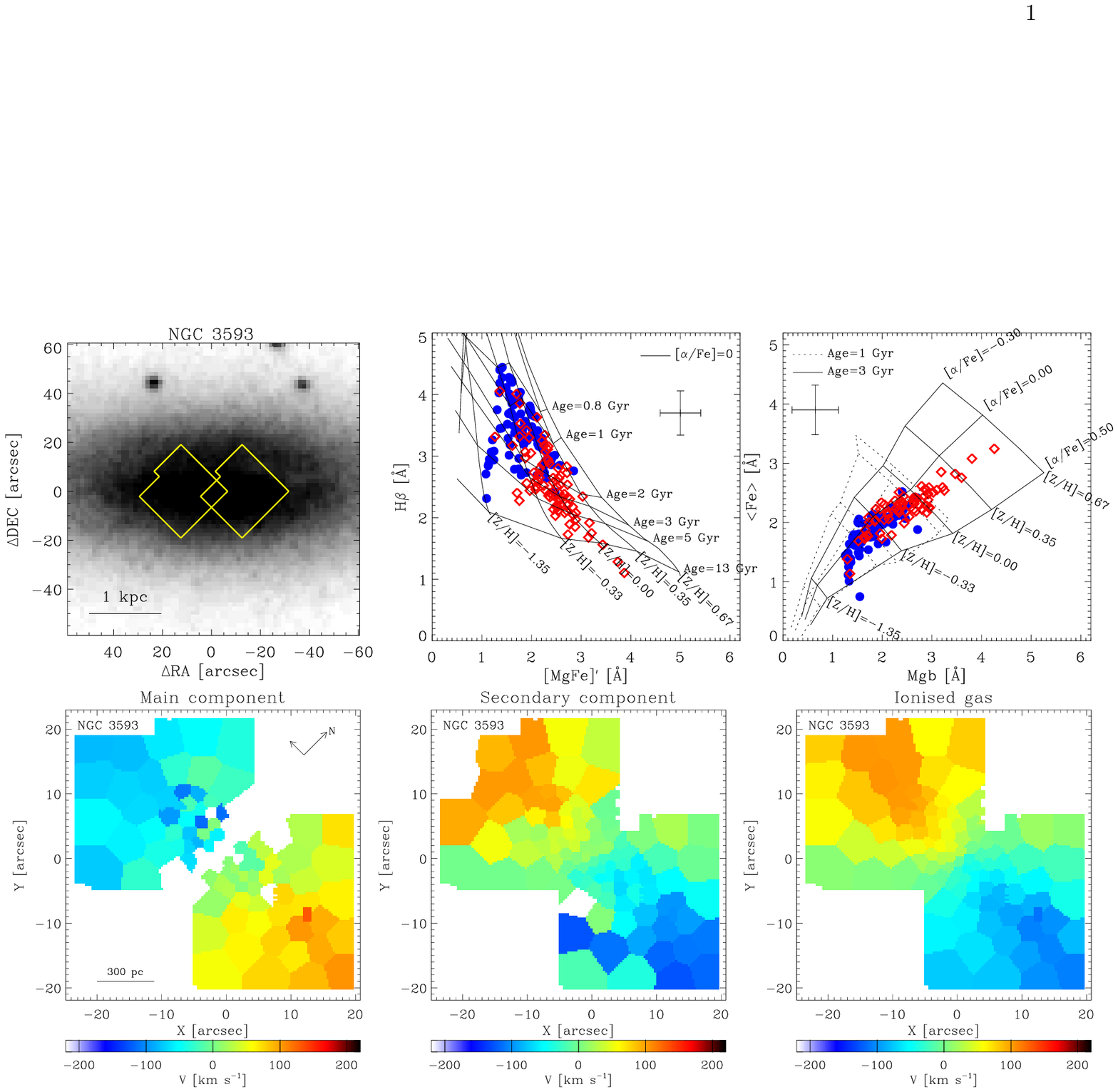}
\caption{Results of the spectroscopic decomposition in NGC~3593. Top
  panels: Observed field of view, absorption line indices, and
  prediction from stellar population models \citep{Thomas+11}; red and
  blue symbols refer to the main and secondary stellar components,
  respectively. Bottom panels: 2-dimensional velocity fields of the
  main and secondary stellar components, and of the ionized gas
  component.}
\label{fig:003_1}
\end{figure}

\section{Analysis: Spectroscopic Decomposition}

We observed with the Visible Multi Object Spectrograph (VIMOS)
integral-field unit at the Very Large Telescope (VLT) three spiral
galaxies, which are known to host counter-rotating stellar disks:
NGC~3593 \citep{Bertola+96}, NGC~4550 \citep{Rubin+92}, and NGC~5197
\citep{Vergani+07}. Our spectroscopic decomposition technique exploits
the differences in kinematics and stellar populations of the two
components to separate their contribution from the combined observed
spectrum. At each position on the galaxy, the code builds two
synthetic templates (one for each stellar component) as linear
combination of spectra from an input stellar library, and convolves
them with two Gaussian line-of-sight velocity distributions with
different kinematics. Gaussian functions are also added to the
convolved synthetic templates to account for ionized-gas emission
lines (H$\gamma$, H$\beta$, [O {\small III}], and [N {\small I}]). The
spectroscopic decomposition code returns the spectra of two best-fit
stellar templates (that represent the two stellar disks) and
ionized-gas emissions, along with the best-fitting parameters of
luminosity of each component, velocity, and velocity dispersion. The
line strength of the Lick indices of the two counter-rotating
components are measured on the two best-fit stellar templates. The
line strength of the absorption line indices are then used to infer
the age, metallicity, abundance ratios, stellar mass-to-light ratio,
and the stellar mass of the two stellar components.

\begin{figure}[t!]
\centering
\includegraphics[width=13cm, clip=]{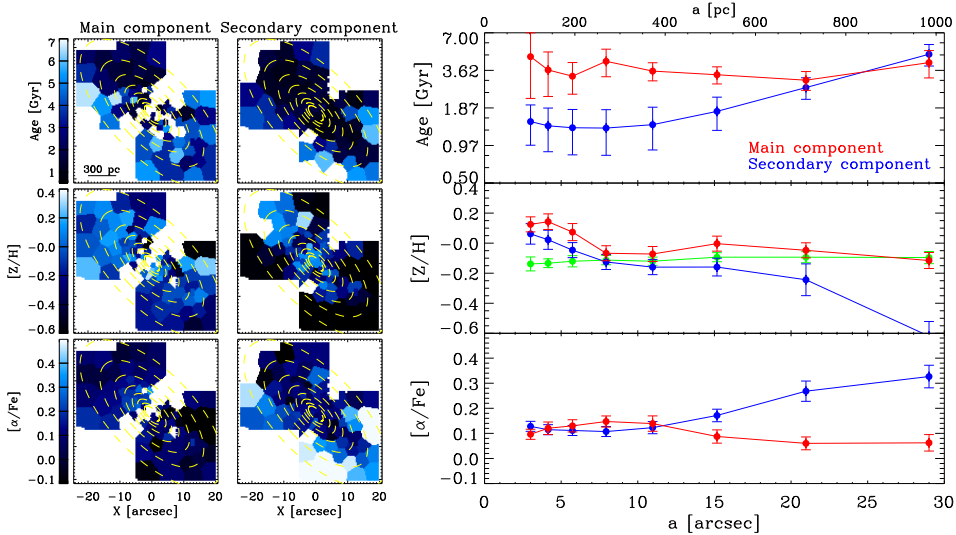}
\caption{Stellar population parameters in NGC~3593. Left panels: Maps
  of the age, metallicity, and $\alpha$-enhancement of the two stellar
  counter-rotating components, as derived by fitting the
  \citet{Thomas+11} models to the measured Lick indices.  Right
  panels: Radial profiles of the total luminosity-weighted age,
  metallicity, and $\alpha$-enhancement within concentric ellipses.}
\label{fig:003_2}
\end{figure}

\section{Results of the Spectroscopic Decomposition}

We successfully separated the contribution of the ionized gas and the
two stellar components from the observed spectrum, and confirm the
presence of two counter-rotating stellar disks in NGC~3593. More
details on this galaxy and other galaxies in our sample are in
\citet{Coccato+11, Coccato+13b}.

\begin{figure}[t!]
\centering
\includegraphics[width=13cm, bb=88 357 980 680, clip=]{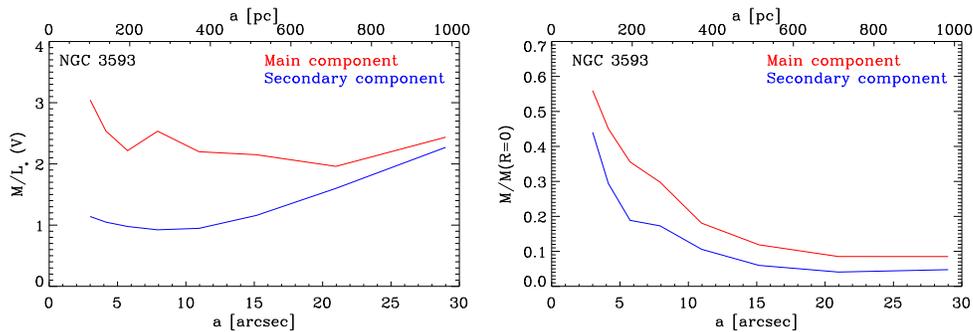}
\caption{Left panel: Radial profiles of the mass-to-light ratio in the
  $V$ band for NGC~3593.  Right panel: Mass surface density profiles
  relative to the central value for NGC~3593.}
\label{fig:003_3}
\end{figure}

Fig.~\ref{fig:003_1} shows the observed field of view, the
2-dimensional velocity fields of the stellar and ionized gas
components, and the line strength of the absorption line indices
(H$\beta$, Mg~{\it b}, [MgFe], and $\langle$Fe$\rangle$).  
Fig.~\ref{fig:003_2} shows the 2-dimensional maps of the age, metallicity
and $\alpha$-enhancement, and their radial profiles. 
Fig.~\ref{fig:003_3} shows the radial profiles of the stellar mass-to-light
ratio and the mass of the two components.

The secondary disk is on average less luminous and less massive than
the main stellar disk, and it rotates along the same direction as the
ionized gas component. The two counter-rotating stellar components
have different stellar populations. The secondary stellar disk is
younger, more metal poor, and more $\alpha$-enhanced than the main
galaxy stellar disk.  The age difference between the two components is
higher within 500 pc: the main component is $(3.7 \pm 0.6)$ Gyr old,
whereas the secondary component is $(1.4 \pm 0.2)$ Gyr old. From the
difference in age of the two components, we are able to date the
formation of the counter-rotating stellar disk to $\sim$2 Gyr ago,
i.e. $\sim$1.6 Gyr after the formation of the main galaxy disk.  Both
components have high metallicity in the central 150 pc followed by a
declining profile. The main component is on average more metal rich
($[Z/{\rm H}]_1=-0.04\pm0.03$) than the secondary component ($[Z/{\rm
    H}]_2=-0.15 \pm0.07$).

\section{Conclusions}

Our results rule out internal processes, such as bar dissolution,
since they would lead to the formation of two counter-rotating disks
with the same properties.

External acquisition of gas followed by star formation, or mergers of
two galaxies with similar mass are both viable mechanisms. The
accretion scenario predicts that the secondary component is the
younger, and that the two components have different metallicity, as they
are formed from different gas clouds. In the second scenario, the
properties of the two stellar component depend on the properties of
the merging systems and star formation episodes triggered by the
merger.

A larger sample is therefore required to statistically determine
which mechanism is the most efficient to build counter-rotating
stellar disks.


\end{document}